\documentstyle[aps,eqsecnum,preprint,tighten,floats]{revtex}
\begin{document}
\preprint{WUGRAV-95-2}
\draft
\title{Gravitational waves from inspiralling compact binaries: \\
       Parameter estimation using second-post-Newtonian waveforms}
\author{Eric Poisson and Clifford M.~Will}
\address{McDonnell Center for the Space Sciences,
         Department of Physics, \\ Washington University,
         St.~Louis, Missouri 63130}
\maketitle
\begin{abstract}
The parameters of inspiralling compact binaries can be estimated using
matched filtering of gravitational-waveform templates against the
output of laser-interferometric gravitational-wave detectors.  The
estimates are most sensitive to the accuracy with which the phases
of the template and signal waveforms match over the many cycles
received in the detector frequency bandwidth. Using a recently
calculated formula, accurate to second post-Newtonian (2PN) order
[order $(v/c)^4$, where $v$ is the orbital velocity], for the
frequency sweep ($dF/dt$) induced by gravitational radiation damping,
we study the statistical errors in the determination
of such source parameters as the ``chirp mass'' $\cal M$, reduced
mass $\mu$, and spin parameters $\beta$ and $\sigma$ (related to
spin-orbit and spin-spin effects, respectively).  We find that
previous results using template phasing accurate to 1.5PN order
actually underestimated the errors in $\cal M$, $\mu$, and $\beta$.
Templates with 2PN phasing yield somewhat larger measurement errors
because the 2PN corrections act to suppress slightly the importance
of spin-orbit contributions to the phase, thereby increasing the
measurement error on $\beta$. This, in turn, results in larger
measurement errors on $\cal M$ and $\mu$ because of the strong
correlations among the parameters.  For two inspiralling neutron
stars, the measurement errors increase by less than 16 percent.
\end{abstract}
\section{Introduction}

Inspiralling compact binary systems, composed of neutron stars
and/or black holes, have been identified \cite{Thorne1987,Schutz}
as the most promising
source of gravitational waves for interferometric detectors
such as the American LIGO (Laser Interferometer
Gravitational-wave Observatory \cite{LIGO})
and the European Virgo \cite{Virgo}.
These systems evolve under the influence of gravitational
radiation reaction, so that the gravitational-wave signal increases
in amplitude as its frequency sweeps through the detector frequency
bandwidth, from approximately 10 Hz to 1000 Hz. (This characteristic
signal is often referred to as a ``chirp.'') Inspiralling compact
binaries are especially promising because
recent estimates \cite{Narayan,Phinney}
indicate that their event rate could be as large
as one hundred per year, for signals detectable out to hundreds of
Mpc by the advanced version of LIGO, and because the signal can be
accurately predicted using general relativity.

That the signal can be calculated with high accuracy is essential
for the measurement of the source parameters \cite{Cutleretal},
which include
distance, position in the sky, orientation of the orbital
plane, and the masses and spins of the companions. Loosely
speaking, the measured signal is passed through a linear filter
constructed from the expected signal $h(t;\bbox{\theta})$ and
the spectral density of the detector noise \cite{Wainstein}
(see below). The
signal-to-noise ratio $S/N (\bbox{\theta})$ is then computed.
The expected signal and the signal-to-noise ratio are expressed
as functions of the vector $\bbox{\theta}$ which collectively
represents the source parameters. The {\it actual} value of these
parameters, which we denote $\bbox{\tilde{\theta}}$, is unknown
prior to the measurement. When $\bbox{\theta} =
\bbox{\tilde{\theta}}$ the linear filter becomes the Wiener optimum
filter which is well known to yield the largest possible
signal-to-noise ratio \cite{Wainstein}.
The source parameters can therefore be determined by
maximizing $S/N(\bbox{\theta})$ over a broad collection
of expected signals $h(t;\bbox{\theta})$, loosely referred
to as ``templates.''

The gravitational-wave signal can be characterized by a growing
amplitude and a phase which accumulates nonlinearly
with time \cite{Thorne1987}. The signal undergoes a number $N$
of oscillations, varying from 600
to 16 000 depending on the nature of the system (see below),
as the frequency sweeps through the detector bandwidth.
It has been established that it is the {\it phasing} of the
signal which plays the largest role in parameter estimation
\cite{Cutleretal,FinnChernoff,CutlerFlanagan}.
This is because a slight variation in the
parameters can quickly cause $h(t;\bbox{\theta})$ to get
out of phase with respect to the true signal
$h(t;\bbox{\tilde{\theta}})$, thus seriously reducing
$S/N(\bbox{\theta})$ from its maximum possible value.
Therefore a good match between the template's phase and that of the
measured signal, throughout the $N$ cycles, singles out, to a
large extent, the value of the source parameters.  (Clearly, this
is only true for those parameters which affect the phasing of the
waves, such as the masses and spins of the companions; see below.)

In principle, the gravitational-wave signal from an inspiralling
compact binary can be calculated exactly using general relativity
(this would require the numerical integration of Einstein's
equations). In practice, however, one must rely on some
approximation scheme. It appears appropriate, in this context,
to adopt a slow-motion approximation \cite{Will},
and to solve the field equations using a combination
of post-Newtonian and post-Minkowskian expansions
\cite{BlanchetDamour}. To date, the
waveform has been calculated accurately through order $(v/c)^4$
(where $v$ is the orbital velocity)
beyond the leading-order, quadrupole-formula expression
\cite{BDIWW,BDI,WW}. Leading-order expressions are
referred to as ``Newtonian;'' the waveform is
therefore known to second-post-Newtonian, or 2PN, order.

The detailed expression for the 2PN waveform is complicated:
the dependence on the various angles (position of the source
in the sky, orientation of the detector, orientation of the
polarization axes) is not simple, and the waves have several
frequency components given by the harmonics of the
orbital frequency (assuming that the orbit is
circular \cite{Peters,LincolnWill}).
For the purpose of this paper, and
following Cutler and Flanagan \cite{CutlerFlanagan},
we shall use a simplified
expression for the waveform. We shall ignore all post-Newtonian
corrections to the wave's {\it amplitude}, and single out its
dominant frequency component at twice the orbital frequency
\cite{Thorne1987}. Thus, setting $G=c=1$,
\begin{equation}
h(t;\bbox{\theta}) = r^{-1} Q(\mbox{angles})
{\cal M} (\pi {\cal M} F)^{2/3} \cos \Phi(t).
\label{1.1}
\end{equation}
Here, $r$ is the distance to the source, $Q$ a function of
the various angles mentioned above,
$F(t)$ the gravitational-wave
frequency, and $\Phi(t) = \int 2\pi F(t) dt$ the phase. We have
also introduced the (so-called) {\it chirp mass} $\cal M$: If
$\mu = m_1 m_2 / (m_1+m_2)$ is the reduced mass and $M=m_1+m_2$
the total mass, then
\begin{equation}
{\cal M} = \eta^{3/5} M, \qquad
\eta = \mu/M.
\label{1.2}
\end{equation}

In Eq.~(\ref{1.1}) we use the most accurate expression available
for the phase function $\Phi(t)$. It is determined by the 2PN
expression for the frequency sweep \cite{BDIWW},
\begin{eqnarray}
\frac{dF}{dt} &=& \frac{96}{5\pi {\cal M}^2} (\pi {\cal M} F)^{11/3}
\biggl[ 1 - \biggl( \frac{743}{336} + \frac{11}{4} \eta \biggr)
(\pi M F)^{2/3} + (4 \pi - \beta) (\pi M F)
\nonumber \\ & & \mbox{}
+ \biggl( \frac{34103}{18144} + \frac{13661}{2016}\eta +
\frac{59}{18} \eta^2 + \sigma \biggr) (\pi M F)^{4/3} \biggr].
\label{1.3}
\end{eqnarray}
Apart from the parameters introduced previously, $dF/dt$ also
depends on $\bbox{\hat{L}}$, the direction of orbital angular
momentum, and on $\bbox{S}_1$ and $\bbox{S}_2$, the spin angular
momentum of each companion. This dependence is hidden in the
``spin-orbit'' parameter \cite{KWW}
\begin{equation}
\beta = \frac{1}{12} \sum_{i=1}^2
\bigl[ 113 (m_i/M)^2 + 75\eta \bigr]
\bbox{\hat{L}} \cdot \bbox{\chi}_i,
\label{1.4}
\end{equation}
where $\bbox{\chi}_i = \bbox{S}_i/{m_i}^2$,
and the ``spin-spin'' parameter \cite{KWW}
\begin{equation}
\sigma = \frac{\eta}{48} \bigl(
-247 \bbox{\chi}_1 \cdot \bbox{\chi}_2 +
721 \bbox{\hat{L}} \cdot \bbox{\chi}_1
\bbox{\hat{L}} \cdot \bbox{\chi}_2 \bigr).
\label{1.5}
\end{equation}

The purpose of this paper is to estimate the {\it anticipated}
accuracy with which the various parameters (such as $\cal M$,
$\eta$, $\beta$, and $\sigma$) can be determined during a
gravitational-wave measurement. This analysis differs from that
of Cutler and Flanagan \cite{CutlerFlanagan}
in that it incorporates terms of 2PN order
into the phasing [terms of order $(\pi M F)^{4/3}$ in
Eq.~(\ref{1.2})]; their calculations were accurate only
through 1.5PN order (terms of order $\pi M F$).
Previous analyses also include
Refs.~\cite{FinnChernoff,KrolakA,KrolakB,KrolakC}.

Our main conclusion is that 1.5PN phasing {\it underestimates}
the uncertainty in such parameters as $\cal M$, $\eta$, and
$\beta$: 2PN phasing predicts somewhat larger measurement
errors. This is true even when no attempt is made to determine
the spin-spin parameter $\sigma$. If, however, $\sigma$ is also
estimated, then the measurement errors become even larger. This
is because the number of estimated parameters has increased
with respect to the number contained in the 1.5PN waveform.

An independent analysis by Cutler and Flanagan \cite{CFprep}
shows that 2PN
waveforms are not sufficiently accurate for the purpose of parameter
estimation: they produce systematic errors which are larger
than the statistical errors inherent to the measurement process.
This is because the 2PN waveform fails to remain in phase with
the true general-relativistic signal, even when the source
parameters are exactly matched
\cite{paperII,TN,TS}. To construct templates such that
the systematic errors will fall below the measurement errors will
require an expression for the wave's phasing accurate
through at least 3PN order \cite{paperII,TN,TS}.
To achieve such a high degree of accuracy is a major challenge
for gravitational-wave theorists.

\begin{table}[t]
\caption[Table I]{Contributions to the accumulated number
of wave cycles measured in a LIGO/Virgo-type detector. The
frequency entering the bandwidth is 10 Hz (seismic limit);
the frequency leaving is 1000 Hz (system A) (shot noise),
and 360 Hz (system B) and 190 Hz (system C) (innermost
circular orbit). The various contributions correspond
to various terms in Eq.~(1.3). Newtonian: first term
within the square brackets; 1PN: second term; tail:
$4\pi (\pi M F)$; spin-orbit: $-\beta (\pi M F)$;
2PN: $(\pi M F)^{4/3}$ terms with $\sigma=0$; and
spin-spin: $\sigma (\pi M F)^{4/3}$.}
\begin{tabular}{lccc}
system & A & B & C \\
\hline
Newtonian & 16 050 & 3 580 & 600 \\
1PN & 439 & 212 & 59 \\
tail & -208 & -180 & -51 \\
spin-orbit & $17\beta$ & $14\beta$ & $4\beta$ \\
2PN & 9 & 10 & 4 \\
spin-spin & $-2\sigma$ & $-3\sigma$ & $-\sigma$
\end{tabular}
\end{table}

We shall consider the following three ``canonical'' binary systems:
\begin{quote}
System A: two neutron stars, with $m_1=m_2=1.4 M_\odot$;

System B: neutron star and black hole, with $m_1=1.4M_\odot$
      (the neutron star) and $m_2=10 M_\odot$ (the black hole); and

System C: two black holes, with $m_1=m_2=10 M_\odot$.
\end{quote}
For each of these systems Table I summarizes the contribution from
each term in Eq.~(\ref{1.3}) to the total number of gravitational-wave
cycles received in a LIGO/Virgo-type detector.

The remainder of the paper is organized as follows. In Sec.~II we
review the theory of parameter estimation, as developed in
previous papers by Finn \cite{Finn} and Cutler and
Flanagan \cite{CutlerFlanagan}. In Sec.~III
we carry out the calculations for the waveform (\ref{1.1}),
and compute the anticipated uncertainty in the
measured values of the source parameters. Our results
are summarized and discussed in Sec.~IV.

\section{Parameter estimation: theory}

The theory of detection and measurement of gravitational-wave
signals was put on a firm statistical foundation, rather
similar to that underlying the theory of radar
detection \cite{Wainstein,Helstrom}. This
was done by various authors, including Finn \cite{Finn}
and Cutler and Flanagan \cite{CutlerFlanagan}.
In this section we review the various aspects of
the theory which are relevant for our purposes.

We assume that some criterion has been applied to conclude
that a signal originating from an inspiralling compact binary
has been received by a network of gravitational-wave detectors.
It is therefore known that a signal of the form $h(t;\bbox{\theta})$
has passed through the detectors, and we seek to determine the
value of the source parameters $\bbox{\theta}$ and the
measurement error $\Delta \bbox{\theta} = \bbox{\theta}
- \bbox{\tilde{\theta}}$, where $\bbox{\tilde{\theta}}$ denotes
the true value.

Finn \cite{Finn}
has derived an expression for $p(\bbox{\theta}|s)$,
the probability that the gravitational-wave signal is characterized
by the parameters $\bbox{\theta}$, {\it given} that the detector
output is $s(t)$ {\it and} that a signal $h(t;\bbox{\theta})$ --- for
{\it any} value of the parameters $\bbox{\theta}$ --- is present.
The detector output is given by
\begin{equation}
s(t) = h(t;\bbox{\theta}) + n(t),
\label{2.1}
\end{equation}
where $n(t)$ represents the detector noise, assumed to be a
stationary, Gaussian random process. Finn shows that
\begin{equation}
p(\bbox{\theta}|s) \propto p^{(0)}(\bbox{\theta})
\exp \Bigl[ -{\textstyle \frac{1}{2}}
\Bigl( h(\bbox{\theta}) - s
\Bigm| h(\bbox{\theta}) - s \Bigr) \Bigr],
\label{2.2}
\end{equation}
where $p^{(0)}(\bbox{\theta})$ is the a priori
probability that the signal is characterized by
$\bbox{\theta}$ (this represents our prior information
regarding the possible value of the parameters) and where the
constant of proportionality is independent of $\bbox{\theta}$.

The inner product $(\cdot | \cdot)$ is defined as
follows \cite{CutlerFlanagan}.
The statistical properties of the detector noise can be summarized
by its autocorrelation function $C_n(\tau) = \langle
n(t) n(t+\tau) \rangle$, where $\langle \cdot \rangle$
denotes a time average. (It is assumed that the noise
has zero mean.) The Fourier transform of the autocorrelation
function gives the noise spectral density
\begin{equation}
S_n(f) = 2 \int_{-\infty}^\infty
C_n(\tau) e^{2\pi i f \tau} d\tau,
\label{2.3}
\end{equation}
which is defined for $f>0$ only.
The inner product is defined so that the
probability for the noise $n(t)$ to have a particular
realization $n_0(t)$ is given by $p(n=n_0) \propto
\exp[-(n_0|n_0)/2]$. It is given by
\begin{equation}
(g | h) = 2 \int_0^\infty
\frac{\tilde{g}^*(f) \tilde{h}(f) +
\tilde{g}(f) \tilde{h}^*(f)}{S_n(f)} df,
\label{2.4}
\end{equation}
where $\tilde{g}(f)$ is the Fourier transform of $g(t)$,
\begin{equation}
\tilde{g}(f) = \int_{-\infty}^\infty
g(t) e^{2\pi i f t} dt;
\label{2.5}
\end{equation}
an asterisk denotes complex conjugation.

We define $\rho$, the signal-to-noise ratio associated
with the measurement, to be
the norm of the signal $h(t;\bbox{\theta})$,
\begin{equation}
\rho^2 = (h|h) = 4 \int_0^\infty
\frac{ |\tilde{h}(f)|^2 }{S_n(f)} df,
\label{2.6}
\end{equation}
evaluated at $\bbox{\theta} = \bbox{\tilde{\theta}}$,
the true value of the source parameters. This
is the largest possible value of the
signal-to-noise ratio, for
$\tilde{h}^*(f;\bbox{\tilde{\theta}})/S_n(f)$ is just
the Fourier transform of the Wiener optimum
filter \cite{Wainstein}.

In a given measurement, characterized by the particular
detector output $s(t)$, the true value of the source
parameters can be estimated by locating the value
$\bbox{\hat{\theta}}$ at which the probability
distribution function (\ref{2.2}) is a maximum. This
is the so-called maximum-likelihood
estimator \cite{Wainstein}. In the
limit of large signal-to-noise ratio, to which we
henceforth specialize, $p(\bbox{\theta}|s)$ will be
strongly peaked about that value. We now derive a
simplified expression for $p(\bbox{\theta}|s)$
appropriate for this limiting case.

We first assume that
$p^{(0)}(\bbox{\theta})$ is nearly uniform near
$\bbox{\theta} = \bbox{\hat{\theta}}$. This indicates
that the prior information is practically irrelevant
to the determination of the source parameters; we shall
relax this assumption below. Then, denoting
$\xi(\bbox{\theta}) =
(h(\bbox{\theta}) -s|h(\bbox{\theta})-s)$,
we have that $\xi$ is minimum at $\bbox{\theta} =
\bbox{\hat{\theta}}$. It follows that
this can be expanded as
\begin{equation}
\xi(\bbox{\theta}) = \xi(\bbox{\hat{\theta}}) +
{\textstyle \frac{1}{2}}
\xi_{,ab}(\bbox{\hat{\theta}})
\Delta \theta^a \Delta \theta^b + \cdots,
\label{2.7}
\end{equation}
where ``$,a$'' denotes partial differentiation with
respect to the parameter $\theta^a$, and
$\Delta \theta^a = \theta^a - \hat{\theta}^a$;
summation over repeated indices is understood.
We assume that $\rho$ is sufficiently large that
the higher-order terms can be neglected.
Calculation yields $\xi_{,ab} = (h_{,ab}|h-s)
+ (h_{,a}|h_{,b})$, and we assume once more that
$\rho$ is large enough that the first term can be
neglected (see Cutler and Flanagan
\cite{CutlerFlanagan} for details). We arrive at
\begin{equation}
p(\bbox{\theta}|s) \propto p^{(0)}(\bbox{\theta})
\exp \Bigl[ - {\textstyle \frac{1}{2}}
\Gamma_{ab} \Delta \theta^a \Delta \theta^b \Bigr],
\label{2.8}
\end{equation}
where
\begin{equation}
\Gamma_{ab} = \bigl( h_{,a} \bigm| h_{,b} \bigr),
\label{2.9}
\end{equation}
evaluated at $\bbox{\theta} = \bbox{\hat{\theta}}$,
is the Fisher information matrix \cite{Helstrom}.

We therefore see that in the limit of large
signal-to-noise ratio,
$p(\bbox{\theta}|s)$ takes a Gaussian form.
{}From Eq.~(\ref{2.8}) it can be established that the
variance-covariance matrix $\Sigma^{ab}$ is given by
\begin{equation}
\Sigma^{ab} \equiv \langle \Delta \theta^a
\Delta \theta^b \rangle = (\bbox{\Gamma}^{-1})^{ab}.
\label{2.10}
\end{equation}
Here, $\langle \cdot \rangle$ denotes an average over
the probability distribution function (\ref{2.8}),
and $\bbox{\Gamma}^{-1}$ is the inverse of the Fisher
matrix. We define the measurement error in the
parameter $\theta^a$ to be
\begin{equation}
\sigma_a =
\bigl\langle (\Delta \theta^a)^2 \bigr\rangle^{1/2}
= \sqrt{\Sigma^{aa}}
\label{2.11}
\end{equation}
(no summation over repeated indices), and
the correlation coefficient between parameters
$\theta^a$ and $\theta^b$ as
\begin{equation}
c^{ab} = \frac{\langle \Delta \theta^a
\Delta \theta^b \rangle}{\sigma_a \sigma_b} =
\frac{\Sigma^{ab}}{
\sqrt{\Sigma^{aa} \Sigma^{bb}}};
\label{2.12}
\end{equation}
by definition each $c^{ab}$ must lie in the range
$(-1,1)$.

Cutler and Flanagan \cite{CutlerFlanagan}
have shown that in the limit of large signal-to-noise ratio,
Eq.~(\ref{2.8}) is valid
even when $p^{(0)}(\bbox{\theta})$ is not uniform
near $\bbox{\theta} = \bbox{\hat{\theta}}$. In such cases
the prior information plays an important role in the
determination of the source parameters. The
exponential factor is still peaked at
$\bbox{\theta} = \bbox{\hat{\theta}}$, but
$\bbox{\hat{\theta}}$ no longer represents the
maximum-likelihood estimate, and the full probability
distribution function $p(\bbox{\theta}|s)$ may not
be a Gaussian.

For simplicity, and following Cutler and Flanagan
\cite{CutlerFlanagan}, we
shall restrict attention to cases such that
$p^{(0)}(\bbox{\theta})$ is a Gaussian,
given by
\begin{equation}
p^{(0)} (\bbox{\theta}) \propto
\exp \Bigl[ - {\textstyle \frac{1}{2}}
\Gamma_{ab}^{(0)} \bigl(\theta^a - \bar{\theta}^a\bigr)
\bigl(\theta^b - \bar{\theta}^b\bigr) \Bigl].
\label{2.13}
\end{equation}
Then $p(\bbox{\theta}|s)$ will {\it also} take a Gaussian
form, and the new variance-covariance matrix will
be given by
\begin{equation}
\bbox{\Sigma} = \bigl( \bbox{\Gamma} +
\bbox{\Gamma^{(0)}} \bigr)^{-1}.
\label{2.14}
\end{equation}
It should be noted that in general,
$p(\bbox{\theta}|s)$ will be peaked at a value
$\langle \bbox{\theta} \rangle$ which differs both
from $\bbox{\hat{\theta}}$ and $\bbox{\bar{\theta}}$.

\section{Parameter estimation: calculations}

We proceed with the calculation of the Fisher information
matrix, Eq.~(\ref{2.9}), for gravitational-wave signals
of the form (\ref{1.1}), and for
gravitational-wave detectors of the LIGO/Virgo type.
For such detectors the anticipated noise spectral density
can be approximated by the analytic
expression \cite{CutlerFlanagan}
\begin{equation}
S_n(f) = {\textstyle \frac{1}{5}} S_0
\bigl[ (f_0/f)^4 + 2 + 2(f/f_0)^2 \bigr],
\label{3.1}
\end{equation}
where $S_0$ is a normalization constant irrelevant for
our purposes, and $f_0$ the frequency at which $S_n(f)$
is minimum; we set $f_0 = 70$Hz, which is appropriate
for advanced LIGO sensitivity \cite{LIGO}.
To mimic seismic noise we assume that Eq.~(\ref{3.1})
is valid for $f > 10 {\rm Hz}$ only, and that
$S_n(f) = \infty$ for $f < 10 {\rm Hz}$.

First, we integrate Eq.~(\ref{1.3}) to obtain expressions
for $\Phi(F)$ and $t(F)$, respectively the phase
and time as functions of gravitational-wave frequency.
(Throughout this section we shall distinguish
between $F$, the function of time describing the
frequency sweep, and $f$, the Fourier-transform variable.)
Expanding in powers of $(\pi M F)^{1/3}$ and truncating
all expressions to 2PN order, we obtain
\widetext
\begin{eqnarray}
\Phi(F) &=& \phi_c - \frac{1}{16} (\pi {\cal M} F)^{-5/3}
\biggl[1 + \frac{5}{3} \biggl( \frac{743}{336} +
\frac{11}{4} \eta \biggr) (\pi M F)^{2/3}
- \frac{5}{2} (4\pi - \beta) (\pi M F)
\nonumber \\ & & \mbox{} +
5 \biggl( \frac{3058673}{1016064} + \frac{5429}{1008}\eta
+ \frac{617}{144} \eta^2 - \sigma \biggr) (\pi M F)^{4/3}
\biggr],
\label{3.2}
\end{eqnarray}
where $\phi_c$ is (formally) the value of $\Phi$ at
$F=\infty$, and
\begin{eqnarray}
t(F) &=& t_c - \frac{5}{256} {\cal M} (\pi {\cal M} F)^{-8/3}
\biggl[1 + \frac{4}{3} \biggl( \frac{743}{336} +
\frac{11}{4} \eta \biggr) (\pi M F)^{2/3} -
\frac{8}{5} (4\pi - \beta) (\pi M F)
\nonumber \\ & & \mbox{} +
2 \biggl( \frac{3058673}{1016064} + \frac{5429}{1008}\eta
+ \frac{617}{144} \eta^2 - \sigma \biggr) (\pi M F)^{4/3}
\biggr],
\label{3.3}
\end{eqnarray}
where (formally) $t_c = t(\infty)$. Of course, the
signal cannot be allowed to reach arbitrarily high
frequencies; it must be cut off at a frequency $F=F_i$
corresponding to the end of the inspiral.
We put $\pi M F_i = (M/r_i)^{3/2} = 6^{-3/2}$;
$r_i = 6M$ is the Schwarzschild radius of the innermost
circular orbit for a test mass moving in the gravitational
field of a mass $M$ \cite{foot1}.
\narrowtext

Next, we take the Fourier transform of Eq.~(\ref{1.1}) and
calculate $\tilde{h}(f) = \int h(t) e^{2\pi i f t} dt$.
It is sufficient to estimate $\tilde{h}(f)$
using the stationary phase approximation,
according to which \cite{Jackson}
\begin{equation}
\int g(t) e^{i \phi(t)} dt \simeq
\biggl[ \frac{2\pi i}{\phi''(t_0)} \biggr]^{1/2}
g(t_0) e^{i \phi(t_0)}
\label{3.4}
\end{equation}
if $g(t)$ varies slowly near $t=t_0$ where the phase has
a stationary point: $\phi'(t_0)=0$ (a prime denotes
differentiation with respect to $t$). Substituting
Eqs.~(\ref{1.1}) and (\ref{3.2}) into (\ref{3.4}),
discarding the irrelevant negative-frequency component,
and neglecting all post-Newtonian corrections to the
amplitude of $\tilde{h}(f)$, we obtain
\begin{equation}
\tilde{h}(f) = {\cal A} f^{-7/6} e^{i \psi(f)},
\label{3.5}
\end{equation}
where ${\cal A} \propto {\cal M}^{5/6} Q(\mbox{angles})/r$,
and
\widetext
\begin{eqnarray}
\psi(f) &=& 2\pi f t_c - \phi_c - \frac{\pi}{4} +
\frac{3}{128} (\pi {\cal M} f)^{-5/3}
\biggl[1 + \frac{20}{9} \biggl( \frac{743}{336} +
\frac{11}{4} \eta \biggr) (\pi M f)^{2/3}
\nonumber \\ & & \mbox{}
- 4(4\pi - \beta) (\pi M f) +
10 \epsilon \biggl( \frac{3058673}{1016064} + \frac{5429}{1008}\eta
+ \frac{617}{144} \eta^2 - \sigma \biggr) (\pi M f)^{4/3}
\biggr].
\label{3.6}
\end{eqnarray}
We have introduced the parameter $\epsilon \equiv 1$. This gives
us the freedom, for future use, of removing the 2PN terms from
$\psi(f)$ by setting $\epsilon=0$.
\narrowtext

We now substitute Eq.~(\ref{3.5}) into (\ref{2.6}) and
calculate the signal-to-noise ratio. We readily obtain
\begin{equation}
\rho^2 = 20 {\cal A}^2 {S_0}^{-1} {f_0}^{-4/3} I(7),
\label{3.7}
\end{equation}
where the integrals $I(q)$ represent various moments of
the noise spectral density:
\begin{equation}
I(q) \equiv \int_{1/7}^{x_i}
\frac{x^{-q/3}}{x^{-4} + 2 + 2x^2} dx,
\label{3.8}
\end{equation}
where $x_i = f_i/f_0 = (6^{3/2} \pi M f_0)^{-1}$ is the
frequency cutoff.

As the next step toward the computation of the Fisher matrix,
we calculate the derivatives of $\tilde{h}(f)$ with respect to
the seven parameters
\begin{equation}
\bbox{\theta} = (\ln{\cal A},f_0 t_c, \phi_c, \ln {\cal M},
\ln \eta, \beta, \sigma).
\label{3.9}
\end{equation}
We obtain
\begin{eqnarray}
\tilde{h}_{,1} &=& \tilde{h}, \nonumber \\
\tilde{h}_{,2} &=& 2\pi i (f/f_0) \tilde{h}, \nonumber \\
\tilde{h}_{,3} &=& -i \tilde{h}, \nonumber \\
\tilde{h}_{,4} &=& - \frac{5 i}{128} (\pi {\cal M} f)^{-5/3}
(1 + A_4 v^2 - B_4 v^3 + C_4 v^4) \tilde{h}, \label{3.10} \\
\tilde{h}_{,5} &=& -\frac{i}{96} (\pi {\cal M} f)^{-5/3}
(A_5 v^2 - B_5 v^3 + C_5 v^4) \tilde{h}, \nonumber \\
\tilde{h}_{,6} &=& \frac{3i}{32} \eta^{-3/5}
(\pi {\cal M} f)^{-2/3} \tilde{h}, \nonumber \\
\tilde{h}_{,7} &=& -\frac{15 i}{64} \eta^{-4/5}
(\pi {\cal M} f)^{-1/3} \tilde{h}, \nonumber
\end{eqnarray}
where $v\equiv(\pi M f)^{1/3}$. We also have defined
\begin{eqnarray}
A_4 &=& \frac{4}{3} \biggl( \frac{743}{336} +
\frac{11}{4} \eta \biggr), \nonumber \\
B_4 &=& \frac{8}{5} (4\pi - \beta),
\label{3.11} \\
C_4 &=& 2 \epsilon \biggl(
\frac{3058673}{1016064} + \frac{5429}{1008} \eta
+ \frac{617}{144} \eta^2 - \sigma \biggr), \nonumber
\end{eqnarray}
and
\begin{eqnarray}
A_5 &=& \frac{743}{168} - \frac{33}{4} \eta, \nonumber \\
B_5 &=& \frac{27}{5} (4\pi - \beta),
\label{3.12} \\
C_5 &=& 18 \epsilon \biggl(
\frac{3058673}{1016064} - \frac{5429}{4032} \eta
- \frac{617}{96} \eta^2 - \sigma \biggr). \nonumber
\end{eqnarray}

Finally, the components of $\bbox{\Gamma}$ can be obtained by
evaluating the inner products $(h_{,a}|h_{,b})$ using
Eq.~(\ref{2.4}). The $\Gamma_{ab}$'s can all be expressed
in terms of the parameters $\bbox{\theta}$, the signal-to-noise
ratio $\rho$, and the integrals $I(q)$. The expressions are too
numerous and lengthy to be displayed here. As illustrating
examples, we quote
\begin{equation}
\Gamma_{1a} = \delta_{1a} \rho^2,
\label{3.13}
\end{equation}
and
\begin{eqnarray}
\Gamma_{46} &=& -\frac{15}{4096} \eta^{-3/5}
(\pi {\cal M} f_0)^{-7/3} \bigl[ J(14) + A_4 J(12)
\nonumber \\ & & \mbox{} -
B_4 J(11) + C_4 J(10) \bigr] \rho^2,
\label{3.14}
\end{eqnarray}
where $J(q) \equiv I(q)/I(7)$. We note that even though the
$\tilde{h}_{,a}$'s are expressed as truncated post-Newtonian
expansions in Eq.~(\ref{3.10}), they must be treated as
{\it exact} when computing $\bbox{\Gamma}$. This is to
ensure that the eigenvalues of the Fisher matrix are always
positive definite.

The variance-covariance matrix $\Sigma^{ab}$ can now be
obtained from Eq.~(\ref{2.14}), and the measurement errors
and correlation coefficients computed from Eqs.~(\ref{2.11})
and (\ref{2.12}). Before doing so, however, we must first
state our assumptions regarding the prior information available
on the source parameters.

We take advantage of the fact that the
dimensionless spin parameters, $\bbox{\chi}_1$ and $\bbox{\chi}_2$,
must necessarily be smaller than unity. (This upper bound is
strict for black holes, but only approximate for neutron stars.)
It follows from Eqs.~(\ref{1.4}) and (\ref{1.5})
that $|\beta|$ must be smaller than
approximately 8.5, and that $|\sigma|$ must be smaller than
approximately 5.0. Following Cutler and
Flanagan \cite{CutlerFlanagan}, we crudely
incorporate this information into our calculations by taking
\begin{equation}
p^{(0)} (\bbox{\theta}) \propto
\exp \bigl[ -{\textstyle \frac{1}{2}} (\beta/8.5)^2
 - {\textstyle \frac{1}{2}} (\sigma/5.0)^2 \bigr].
\label{3.15}
\end{equation}
We consider all other parameters to be
unconstrained \cite{foot2}.

\section{Results and discussion}

Equation (\ref{3.12}) implies that the Fisher matrix is
block diagonal. The parameter $\theta^1 = \ln \cal A$
is therefore entirely uncorrelated with the other parameters,
and we find $\sigma_1 = \Delta {\cal A} / {\cal A} =
1/\rho$, $c^{1a}=0$, in all cases. We shall
no longer be concerned with this parameter.

The results concerning the other parameters are displayed in
Tables II and III. All calculations were carried out assuming
$\rho=10$, and that the companions are spinless, so that
$\beta=\sigma=0$.

\begin{table}[t]
\caption[Table II]{Measurement errors and correlation
coefficients for the estimation of six parameters
($\sigma$ is not estimated), assuming $\rho=10$ and
$\beta=0$. The first column indicates whether or not
prior information was included in the
calculation. The second column gives the value of
$\epsilon$, introduced in Eq.~(3.6); $\epsilon=1$
represents 2PN phasing. Then follows,
in more suggestive notation, $\sigma_2/f_0$ (in
msec), $\sigma_3$ (in radians), $\sigma_4$,
$\sigma_5$, $\sigma_6$, $c^{45}$, $c^{46}$,
and $c^{56}$ (all dimensionless).}
\begin{tabular}{lrrrrrrrrr}
prior & $\epsilon$ & $\Delta t_c$ & $\Delta \phi_c$ &
$\Delta {\cal M} / {\cal M}$ & $\Delta \eta / \eta$ &
$\Delta \beta$ & $c^{{\cal M} \eta}$ &
$c^{{\cal M} \beta}$ & $c^{\eta\beta}$ \\
\hline
\multicolumn{10}{l}{System A (two neutron stars):} \\
\multicolumn{10}{l}{ } \\
yes & 1 & 1.07 & 2.94 & 0.036 \% & 0.279 & 1.33 &
-0.989 & 0.994 & -0.999 \\
no & 1 & 1.08 & 2.97 & 0.037 \% & 0.282 & 1.35 &
-0.989 & 0.994 & -0.999 \\
no & 0 & 1.13 & 4.09 & 0.034 \% & 0.243 & 1.24 &
-0.988 & 0.993 & -0.999 \\
no & -1 & 1.16 & 4.96 & 0.032 \% & 0.213 & 1.15 &
-0.986 & 0.992 & -0.999 \\
\hline
\multicolumn{10}{l}{System B (neutron star and
black hole):} \\
\multicolumn{10}{l}{ } \\
yes & 1 & 1.72 & 2.27 & 0.218 \% & 0.503 & 2.29 &
-0.993 & 0.996 & -0.999 \\
no & 1 & 1.76 & 2.32 & 0.226 \% & 0.523 & 2.38 &
-0.993 & 0.996 & -0.999 \\
no & 0 & 2.04 & 6.24 & 0.191 \% & 0.386 & 1.99 &
-0.990 & 0.994 & -0.999 \\
no & -1 & 2.20 & 8.68 & 0.171 \% & 0.306 & 1.76 &
-0.988 & 0.993 & -0.999 \\
\hline
\multicolumn{10}{l}{System C (two black holes):} \\
\multicolumn{10}{l}{ } \\
yes & 1 & 1.50 & 2.19 & 0.54 \% & 1.46 & 8.19 &
-0.946 & 0.956 & -0.999 \\
no & 1 & 2.40 & 4.99 & 1.96 \% & 5.50 & 30.8 &
-0.996 & 0.997 & -0.999 \\
no & 0 & 3.53 & 9.27 & 1.42 \% & 3.16 & 19.5 &
-0.992 & 0.994 & -0.999 \\
no & -1 & 4.01 & 14.7 & 1.21 \% & 2.22 & 14.9 &
-0.989 & 0.992 & -0.999
\end{tabular}
\end{table}

To obtain the results of Table II we have estimated only
six (including ${\cal A}$) of the seven parameters, leaving
$\sigma$ out. In effect, we have truncated the original Fisher
matrix to a smaller, $6\times 6$, matrix. This amounts to assuming
before measurement that the spin-spin parameter must be very small;
equivalently, this assumption can be implemented by taking
$p^{(0)}(\sigma)$ to be very strongly peaked at $\sigma=0$.

For each of the three canonical systems, the first line
of Table II displays the measurement errors and correlation
coefficients as calculated using 2PN phasing
($\epsilon = 1$) and the prior probability distribution
function (\ref{3.15}). The second line shows the same quantities
calculated without utilizing the prior information. We notice that
the prior information makes virtually no difference for systems A
and B, but is very significant for system C.

The third line of Table II displays the measurement errors and
correlation coefficients assuming no prior information and
1.5PN phasing ($\epsilon=0$). Our values agree with those of
Cutler and Flanagan \cite{CutlerFlanagan,foot3}.
We notice that the errors calculated
using 1.5PN phasing are always {\it larger} for $t_c$ and $\phi_c$,
and {\it smaller} for $\cal M$, $\eta$, and $\beta$, than those
calculated using 2PN phasing. Thus, the measurement errors
on the masses and spins are {\it underestimated} when
evaluated using the less accurate 1.5PN phasing.

This can be explained with a simple argument.
In Eq.~(\ref{3.6}), the 1PN and 2PN terms [of order
$(\pi M f)^{2/3}$ and $(\pi M f)^{4/3}$ respectively] combine,
when $\epsilon=1$, so as to reduce the relative importance of the
$\pi M f$ term, when compared to the situation when
$\epsilon=0$. In other words, the relative contribution
to the total number of wave cycles coming from the $\pi M f$
term is less for 2PN phasing than it is for 1.5PN phasing
(see Table I). It is therefore expected that
2PN phasing will produce larger measurement errors for
$\beta$, since all information about $\beta$
comes from the $\pi M f$ term.
But because $\beta$ is strongly correlated with both
$\cal M$ and $\eta$, it follows that these parameters will
{\it also} come with larger measurement errors. This is indeed
what is observed. It is amusing to test this explanation by
artificially setting $\epsilon=-1$ in our calculations, which
we do in the fourth line of Table II. The argument suggests
that the errors in $\cal M$, $\eta$, and $\beta$ should all
decrease with respect to the values calculated using 1.5PN
phasing, since the relative importance of the $\pi M f$ term
is now increased. This is indeed what the results show.

\begin{table}[t]
\caption[Table III]{Measurement errors and correlation
coefficients for the estimation of all seven parameters,
assuming $\rho=10$, $\beta=0$, and $\sigma=0$. The first
column indicates whether or not prior information was
included in the calculation. The notation is similar to
that of Table II, and $\epsilon=1$ in all cases.}
\begin{tabular}{lrrrrrrrrrrrr}
prior & $\Delta t_c$ & $\Delta \phi_c$ &
$\Delta {\cal M} / {\cal M}$ & $\Delta \eta / \eta$ &
$\Delta \beta$ & $\Delta \sigma$ &
$c^{{\cal M} \eta}$ & $c^{{\cal M} \beta}$ &
$c^{{\cal M} \sigma}$ & $c^{\eta\beta}$ &
$c^{\eta \sigma}$ & $c^{\beta\sigma}$ \\
\hline
\multicolumn{13}{l}{System A (two neutron stars):} \\
\multicolumn{13}{l}{ } \\
yes & 1.28 & 13.3 & 0.047 \% & 0.507 & 1.77 & 4.79 &
-0.956 & 0.996 & -0.648 & -0.964 & 0.835 & -0.660 \\
no & 2.72 & 46.9 & 0.120 \% & 1.578 & 4.53 & 17.3 &
-0.991 & 0.999 & -0.952 & -0.993 & 0.984 & -0.955 \\
\hline
\multicolumn{13}{l}{System B (neutron star and
black hole):} \\
\multicolumn{13}{l}{ } \\
yes & 2.54 & 23.6 & 0.280 \% & 0.873 & 3.02 & 4.74 &
-0.959 & 0.997 & -0.630 & -0.969 & 0.817 & -0.650 \\
no & 7.52 & 95.9 & 0.813 \% & 3.10 & 8.98 & 19.4 &
-0.993 & 0.999 & -0.961 & 0.995 & 0.986 & -0.964 \\
\hline
\multicolumn{13}{l}{System C (two black holes):} \\
\multicolumn{13}{l}{ } \\
yes & 2.22 & 10.4 & 0.55 \% & 1.51 & 8.22 & 4.81 &
-0.849 & 0.920 & 0.191 & -0.984 & 0.257 & -0.081 \\
no & 17.0 & 179 & 7.23 \% & 30.7 & 149 & 74.6 &
-0.995 & 0.998 & -0.962 & -0.999 & 0.984 & -0.978
\end{tabular}
\end{table}

To obtain the results of Table III we have estimated all
seven parameters, including both $\cal A$ and $\sigma$,
and used 2PN phasing. For each of the three systems,
the first line of Table III
displays the measurement errors and correlation coefficients
calculated using the prior probability distribution function
(\ref{3.15}). We notice that the measurement errors are all
significantly larger than those displayed in Table II; this
is expected from the fact that we are now estimating a larger
number of parameters.

It is interesting to ask how the measurement errors
increase as the number of estimated parameters increases. In
Ref.~\cite{CutlerFlanagan}, Cutler and Flanagan initially
estimate only five of their six parameters,
leaving $\beta$ out. When
they next include $\beta$ in their calculations, they find
that the measurement errors on $\cal M$ and $\mu$ increase
by a factor of order 10. In this paper, on the other hand,
we have initially estimated only six of our seven parameters,
leaving $\sigma$ out. When
we next include $\sigma$ in our calculations, we find that
the measurement errors on $\cal M$, $\mu$, and $\beta$ only
increase by a factor of order unity. Thus, the inclusion of
$\sigma$ in the calculation has less dramatic consequences than
the inclusion of $\beta$. This confirms a conjecture formulated
by Cutler and Flanagan \cite{CutlerFlanagan} at the end of
their Sec.~III. That this is so is largely due to the importance
of prior information in the estimation of $\sigma$.

In the second line of Table III we display the results
obtained when the prior information is {\it not} included
in the calculations. We notice that for all systems, the
prior information indeed plays a very important role. In fact,
we see that Eq.~(\ref{3.15}) provides nearly {\it all}
of the information regarding the spin-spin parameter
$\sigma$. This explains why the measurement error on $\sigma$
is always nearly equal to 5.0, and its correlation coefficient
with other parameters significantly smaller than unity.
Of course, these results only apply to gravitational-wave
measurements with $\rho=10$. To bring the error on $\sigma$
well below the a priori constraint $\sigma < |5.0|$,
say $\Delta \sigma \lesssim 3$, the measurement
would require a signal-to-noise ratio larger than approximately
45 for system A, 50 for system B, and 110 for system C.

We conclude with the following remark. It is clear that the
results displayed in Tables II and III depend on a fairly large
number of simplifying assumptions, and that a more careful
treatment might produce somewhat different numbers. These
assumptions include: (i) the simplified form (\ref{1.1}) for
the waveform; (ii) the neglect of (not yet calculated)
higher-order terms in the post-Newtonian expansion (\ref{1.3});
(iii) the neglect of $O(1\rho)$ corrections in the expression
(\ref{2.8}), (\ref{2.9}) for $p(\bbox{\theta}|s)$; (iv)
the analytic model (\ref{3.1}) for the noise spectral
density; and (v) our rather crude incorporation of the prior
information. We shall leave for future work the difficult task
of carefully examining the effect of these assumptions on
our results.

\section*{Acknowledgments}

Our warmest thanks to Eanna Flanagan for many useful conversations
and his detailed comments on the manuscript. This work was
supported by the National Science Foundation
under Grant No.~PHY 92-22902 and the National Aeronautics and
Space Administration under Grant No.~NAGW 3874.


\begin{references}
\bibitem{foot2} Our calculations should take into account the
        fact that $\eta$ is restricted to the interval $(0,1/4]$.
        However, this information cannot be incorporated using
        a Gaussian probability distribution function such as
        Eq.~(3.15). To keep the analysis simple we have preferred
        to leave $\eta$ unrestricted.
\bibitem{foot3} Cutler and Flanagan use $\ln\mu$ as a parameter
        instead of $\ln\eta$. It is easy to show, by performing
        the coordinate transformation, that $\Delta \mu/ \mu =
        (2/5) \Delta \eta/\eta$ and $c^{\mu a} = c^{\eta a}$.
\end{references}
\end{document}